\documentclass[pra, twocolumn, superscriptaddress, aps, showpacs, amsmath, amssymb,  nofootinbib]{revtex4}

\usepackage{graphicx}
\usepackage{dcolumn}
\usepackage{bm}
\usepackage{bbm}
\usepackage{psfrag}

\newcommand{\im}{\mathbbm{i}}

\newcommand{\ket}[1]{\vert #1 \rangle}

\newcommand{\meanv}[1]{\big\langle #1 \big\rangle}
\newcommand{\var}[1]{\left( \Delta #1\right)^2}
\newcommand{\varsb}[1]{\left[ \Delta #1\right]^2}

\begin{document}

\title{Manipulating mesoscopic multipartite entanglement with atom-light interfaces}
\date{\today}

\author{J. Stasi\'nska}
\affiliation{Grup de F\'isica Te\`orica, Universitat
Aut\`onoma de Barcelona, 08193 Bellaterra (Barcelona), Spain.}

\author{C. Rod\'o}
\affiliation{Grup de F\'isica Te\`orica, Universitat
Aut\`onoma de Barcelona, 08193 Bellaterra (Barcelona), Spain.}

\author{S. Paganelli}
\affiliation{Grup de F\'isica Te\`orica, Universitat
Aut\`onoma de Barcelona, 08193 Bellaterra (Barcelona), Spain.}

\author{G. Birkl}
\affiliation{Institut f\"ur Angewandte Physik, Technische Universit\"at Darmstadt, 64289 Darmstadt, Germany.}

\author{A. Sanpera}
\affiliation{ICREA-Instituci\'o Catalana de Recerca i Estudis Avan\c cats, Llu\'is Companys 23, 08010 Barcelona, Spain.}
\affiliation{Grup de F\'isica Te\`orica, Universitat
Aut\`onoma de Barcelona, 08193 Bellaterra (Barcelona), Spain.}

\begin{abstract}
Entanglement between two macroscopic atomic ensembles induced by measurement on an ancillary light system has proven to be a powerful method for engineering quantum memories and quantum state transfer. Here we investigate the feasibility of such methods for generation, manipulation and detection of genuine multipartite entanglement (Greenberg-Horne-Zeilinger and clusterlike states) between mesoscopic atomic ensembles without the need of individual addressing of the samples. Our results extend in a non trivial way the Einstein-Podolsky-Rosen entanglement between two macroscopic gas samples reported experimentally in [B. Julsgaard, A. Kozhekin, and E. Polzik, Nature (London) {\bf 413}, 400 (2001)]. We find that under realistic conditions, a second orthogonal light pulse interacting with the atomic samples, can modify and even reverse the entangling action of the first one leaving the samples in a separable state.
\end{abstract}


\maketitle


\section{Introduction}\label{intro}
Matter-light quantum interfaces refer to those interactions that lead to a faithful transfer of correlations between atoms and photons.
The interface, if appropriately tailored, generates an entangled state of light and matter which can be further manipulated (for a review see \cite{2008polzikrev,sherson2006} and references therein).  To this aim, a strong  coupling between atoms and photons is a must. A pioneering method to enhance the coupling is cavity QED, where atoms and photons are made to interact strongly due to the confinement imposed by the boundaries (\cite{harochebook} and references therein). An alternative approach to reach the strong coupling regime in {\it free space}
is to use optically thick atomic samples.

Atomic samples with internal degrees of freedom (collective spin) can be made to interface with light via the Faraday effect,
which refers to the polarization rotation that is experienced by a linearly polarized light propagating inside a magnetic medium.
At the quantum level, the Faraday effect leads to an exchange of fluctuations between light and matter.
As demonstrated by Kuzmich and co-workers \cite{Kuzmich97}, if an atomic sample interacts with a squeezed light whose
polarization is measured afterwards, the collective atomic state is projected into a spin squeezed state (SSS).
Furthermore, to produce a long lived SSS, Kuzmich and coworkers \cite{Kuzmich00} proposed a quantum non demolition (QND) measurement, based on off-resonant light propagating through an atomic polarized sample in its ground state.

A step forward within this scheme is measurement induced entanglement between two macroscopic atomic ensembles. As proposed by Duan {\it et al.} \cite{duan2000b} and demonstrated by Polzik and co-workers \cite{2001Natur.413..400J},
the interaction between a single laser pulse, propagating through two spatially separated atomic ensembles combined with a final projective measurement on the light, leads to an Einstein-Podolski-Rosen (EPR) state of the two atomic ensembles. Due to the QND character of the measurement, the verification of entanglement is done by a homodyne measurement of a second laser pulse that have passed through the samples. From such measurements,
atomic spin variances inequalities can be checked, asserting whether the samples are entangled or not. A complementary
scheme for  measurement induced entanglement  is also introduced in \cite{dilisiPhysRevA,PhysRevA.66.052303}.

The quantum Faraday effect can also be used as a powerful spectroscopic method \cite{SorensenPRL1998}. Tailoring the spatial shape of the light beam, provides furthermore, a detection method with spatial resolution which opens the possibility to detect phases of strongly correlated systems generated with ultracold gases in optical lattices
\cite{demler, ekert-nature,RoscildeNJP2009}.

Here, we analyze the suitability of the Faraday interface in the multipartite scenario. In contrast to the bipartite case, where only one type of entanglement exists, the multipartite case offers a richer situation
\cite{dur-PhysRevLett.83.3562,giedke}. Due to this fact, the verification of entanglement using spin variance inequalities \cite{vanloock-PhysRevA.67.052315} becomes an intricate task.
We address such problem and provide a scheme for the generation and verification of multipartite entanglement between atomic ensembles.
Furthermore, in contrast to the scheme of Julsgaard \cite{2001Natur.413..400J}, where the verification of entanglement requires individual addressing
of the atomic samples, here we eliminate this constraint. Despite the irreversible character of the entanglement induced by measurement, we find that a second pulse can reverse the action of the first one deleting all the entanglement between the atomic samples.
This result has implications in the use of atomic ensembles as quantum memories \cite{Julsgaard2004Nat}.

The paper is organized as follows. In Section \ref{II} we briefly introduce  the interaction Hamiltonian as well as the formalism necessary to proceed towards the main results. In Section \ref{III}, we review the basics of the bipartite measurement induced entanglement  and introduce our scheme to detect entanglement without individual addressing of samples. We explicitly derive the spin variances of the atomic ensembles after measurements. From there, we investigate under which conditions the interaction of the samples with a second field erases all the entanglement created by the first one. In Section \ref{IV}, we tackle the multipartite case where the corresponding detection of spin variance inequalities becomes a much harder problem. Also in this case we show the feasibility of our scheme to generate and detect multipartite entanglement, both Greenberg-Horne-Zeilinger (GHZ) and clusterlike, as well as the generalized conditions for a multipartite entanglement eraser.
In the closing section we summarize our main results and point out few open questions.


\section{Formalism}\label{formalism}\label{II}

The basic concept, underlying the QND atom-light interface we will use, is the dipole interaction between an off-resonant linearly polarized light with a polarized atomic ensemble, followed by a quantum homodyne measurement of light.
On one hand, we consider an ensemble of $N_{\mathrm{at}}$ non interacting alkali atoms with total angular momentum $\mathbf{F}$ prepared in the ground state manifold $\ket{F,m_F}$.  We describe such sample with its collective angular momentum $\hat{\mathbf{J}}=(\hat{J}_x,\hat{J}_y,\hat{J}_z)=\sum_{n=1}^{N_{\mathrm{at}}}\mathbf{F}_n$. Further we assume that all atoms are polarized along the $x$ direction, which corresponds to preparing them in a certain hyperfine state $\ket{F,m_F}$  (e.g. in the case of Cesium the hyperfine ground state  $6S_{1/2}$ with total angular momentum $F=4$ and $m_F=4$). Then, the $\hat{J}_x$ component of the collective spin can be regarded as a classical number $\hat{J}_x \approx \meanv{\hat{J}_x} = N_{\mathrm{at}} F \hbar$, while the orthogonal spin components encode all the quantum character. Due to the above approximation, the orthogonal collective angular momentum components can be treated as canonical conjugate variables, $\left[\hat J_y, \hat J_z \right] = \im \hbar  J_x$.

On the other hand, the polarization of light propagating along the $z$ direction can be described by the Stokes vector $\hat{\mathbf{s}}=(\hat{s}_x,\hat{s}_y,\hat{s}_z)$, whose components correspond to the differences between the number of photons (per time unit) with $x$ and $y$ linear polarizations, $\pm \pi/4$ linear polarizations and the two circular polarizations
\begin{eqnarray}\label{stokes}
\hat{s}_x & = &\frac{\hbar}{2}(\hat{n}_{x}-\hat{n}_{y})\nonumber\\
\hat{s}_y & = &\frac{\hbar}{2}(\hat{n}_{\nearrow}-\hat{n}_{\searrow}) \nonumber\\
\hat{s}_z & = &\frac{\hbar}{2}(\hat{n}_{\circlearrowleft}-\hat{n}_{\circlearrowright}).
\end{eqnarray}
The above operators have dimension of energy. They are convenient for a microscopic description of interaction between light and atoms, however, we will concentrate on the macroscopic variables, defined as $\hat{S}_k(z)=\int_0^{T} \hat{s}_k(z,t)\mathrm{d}t$ $(k=x,y,z)$, where $T$ is the length of the light pulse. Such defined operators correspond now to differences in total number of photons, and obey standard angular momentum commutation rules. For light linearly polarized along the $x$-direction $\hat{S}_x \approx \meanv{\hat{S}_x}= \hbar N_{\mathrm{ph}}/2$. In such case, the orthogonal Stokes components $\hat{S}_y, \hat{S}_z$ fulfill canonical commutation relations and therefore can be treated as conjugated variables.
\begin{figure}
 \centering
 \includegraphics[width=0.3\textwidth]{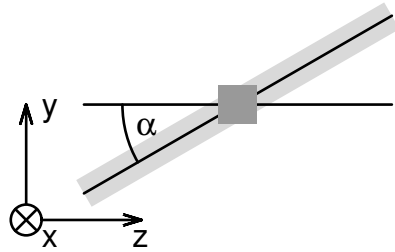}\\
 \caption{A beam passing through an atomic sample at an angle $\alpha$ with respect to $z$.}\label{angle}
\end{figure}

For a light beam propagating through the atomic sample in the $YZ$ plane at a certain angle $\alpha$ with respect to direction $z$ (see Fig. \ref{angle}), the atom-light interaction can be approximated to the following QND effective Hamiltonian,
\begin{equation}\label{ham_alpha}
\hat H^{\mathrm{eff}}_{\mathrm{int}}(\alpha)=-\frac{a}{T} \hat{S}_z (\hat{J}_z \cos\alpha+\hat{J}_y \sin\alpha).
\end{equation}
We have restricted here to the linear coupling between the Stokes operator and the collective atomic spin operator.
The parameter $a=\frac{\gamma}{8 A \Delta} \frac{\lambda^2}{2\pi}$ is a coupling constant
with $A$ being the cross section, $\lambda$ the wave length of light, $\Delta$  the  detuning energy and $\gamma$ the frequency width of atomic excited state.
As one can see from the above expression, the detuning should not be too large for the interaction not to vanish.
For a detailed derivation of such Hamiltonian as well as the conditions under which it is valid, we refer the reader to \cite{julsgaardPHD2003,sherson2006,kupriyanov:032348,2008polzikrev}.
The effective Hamiltonian governs the atomic dynamics (since spin diffusion occurs on a much larger time scale) and the evolution equations are derived straight through the Heisenberg equations for matter and Maxwell-Bloch equations (neglecting retardation effects) for light

\begin{eqnarray}\label{evolution_single}
&& \hat J_y^{\mathrm{out}} = \hat J_y^{\mathrm{in}}-a\hat{S}_z^{\mathrm{in}} J_x \cos{\alpha}\label{Jy}\\
&& \hat J_z^{\mathrm{out}} = \hat J_z^{\mathrm{in}}+ a \hat{S}_z^{\mathrm{in}} J_x \sin{\alpha}\label{Jz}\\
&& \hat S^{\mathrm{out}}_y = \hat S^{\mathrm{in}}_y - a S_x (\hat{J}_z^{\mathrm{in}} \cos\alpha+\hat{J}_y^{\mathrm{in}} \sin\alpha)\label{Sy}\\
&& \hat S^{\mathrm{out}}_z = \hat S^{\mathrm{in}}_z\label{Sz},
\end{eqnarray}
where the operators $\hat{S}_k^{\mathrm{in/out}}=\hat{S}_k(0/L)$ are the Stokes operators characterizing the pulse entering ($z=0$) and leaving ($z=L$) the atomic sample. Analogously, $\hat J_k^{\mathrm{in/out}}$ correspond to initial and final state of atomic spin. From Eq. (\ref{Sy}) it is clear that the polarization of the outgoing light carries information about the collective atomic angular momentum. The quantum character of the interface is reflected at the level of fluctuations, i.e.,
\begin{equation}\label{variation_single}
\var{\hat S^{\mathrm{out}}_y} = \var{\hat S^{\mathrm{in}}_y}
+ a^2 S_x^2 \varsb{(\hat J_z\cos{\alpha}+\hat J_y\sin{\alpha})}.
\end{equation}
At the same time, Eqs. (\ref{Jy}) and (\ref{Jz})  show the QND character of the Hamiltonian, i.e.,  the measured combination $\hat{J}_z^{\mathrm{in}} \cos{\alpha}+\hat{J}_y^{\mathrm{in}}\sin{\alpha}$ is not affected by the interaction since it commutes with the effective Hamiltonian. This fact allows to measure the fluctuations of the atomic spin component with the minimal disruption permitted by Quantum Mechanics.

In the following sections we will generalize the above formalism to the interaction of a light pulse with an arbitrary number, $N_{s}$, of spatially separated atomic samples.  Variables characterizing each sample will be denoted by $\hat{J}_k^{(i)}$, where $i=1,2,\ldots, N_{s} $ denotes the sample and $k=x,y,z$.
In what follows we omit the superscripts $\mathrm{in}$ and $\mathrm{out}$ when they are not necessary.


\section{Bipartite entanglement and entanglement eraser}\label{III}

\begin{figure}
 \centering
 a)\includegraphics[width=0.36\textwidth]{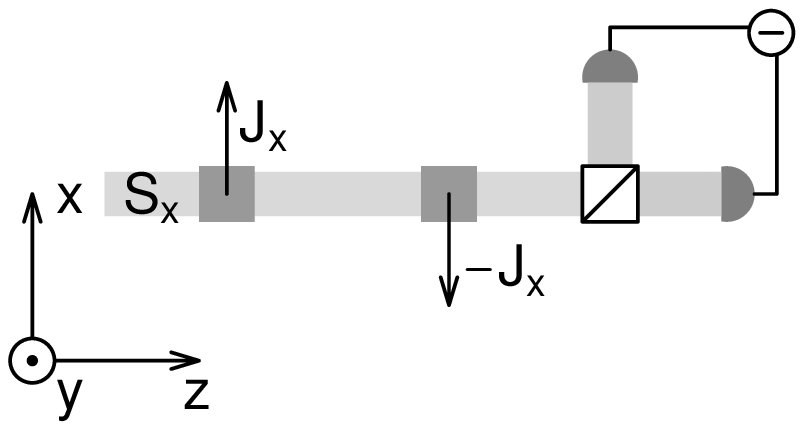}\\
 b)\includegraphics[width=0.36\textwidth]{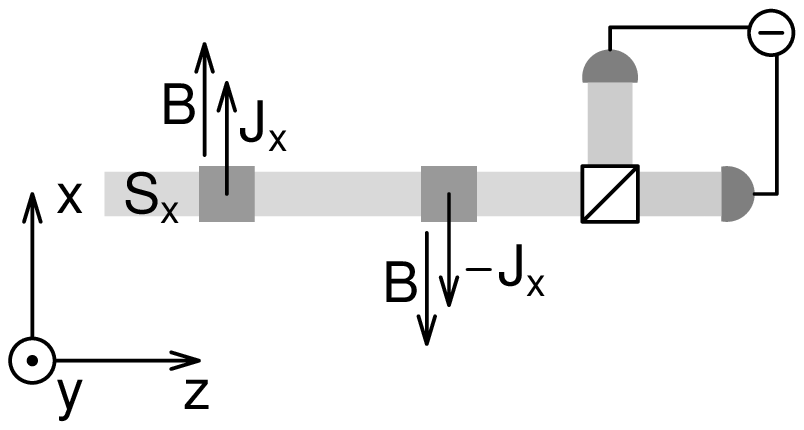}\\
 \caption{Sketch of the experimental setup applied in \cite{2001Natur.413..400J} to generate bipartite entanglement. (a) Entangling pulse. (b) Verifying pulse followed by homodyne measurement. A local magnetic field is added in order to measure two transverse components of the spin with a single light beam.}\label{two_parties_Polzik}
\end{figure}

We begin by briefly reviewing  the atom-light interface scheme implemented in \cite{2001Natur.413..400J}
to entangle two spatially separated atomic samples, as schematically shown in Fig. \ref{two_parties_Polzik}.
In the experimental setup, both
light and atomic samples were strongly polarized along the $x$-direction while light propagated along the $z$-direction. Setting $\alpha=0$ in the effective interaction Hamiltonian,  one can easily derive the equations of motion. The collective polarization of atoms along the $z$-direction is preserved, i.e., ${\partial \hat{J}_z^{(i)}}/{\partial t} = 0$, and Eq. (\ref{Sy}) reads now
\begin{equation}\label{out_bipartite}
\hat{S}_y^{\mathrm{out}}=\hat{S}_y^{\mathrm{in}}-a S_x \left(\hat{J}_z^{(1)}+\hat{J}_z^{(2)}\right).
\end{equation}
Entanglement between the atomic samples is established as soon as the $\hat{S}_y^{\mathrm{out}}$ component of light is measured. Moreover, it
should be emphasized that entanglement is generated independently of the outcome of the measurement. The real challenge, though, is its experimental verification, since spin entanglement criteria rely on spin variances inequalities of
operators of the type $(\hat{J}_y^{(1)}-\hat{J}_y^{(2)})$ and $(\hat{J}_z^{(1)}+\hat{J}_z^{(2)})$. This is so because the maximally entangled EPR state is a coeigenstate of such operators. This fact, in turn, imposes an upper bound on the variances of such operators giving rise to a sufficient and necessary condition for separability \cite{duan2000},
\begin{eqnarray}\label{sep_crit_Duan}
&&\var{\left[|\lambda|\hat{J}_y^{(1)}+\frac{\hat{J}_y^{(2)}}{\lambda}\right]}+\var{\left[|\lambda|\hat{J}_z^{(1)}-\frac{\hat{J}_z^{(2)}}{\lambda}\right]}\nonumber \\
&&\geq\left(\lambda^2+\frac{1}{\lambda^2}\right) \hbar J_x,
\end{eqnarray}
for all $\lambda\in\mathbbm{R}$.

The way to experimentally check \cite{2001Natur.413..400J} the above equation with $\lambda=-1$ was to
add on each sample an external magnetic field, quasi parallel to the $x$-direction (see also
\cite{muschik:062329}). The magnetic field was local, therefore, it did not affect the generation of entanglement. However, it caused a Larmor precession of  the collective atomic momenta, which permitted a simultaneous measurement of the appropriately redefined "canonical variables" $\hat{J}_y^{(1)}+\hat{J}_y^{(2)}$ and $\hat{J}_z^{(1)}+\hat{J}_z^{(2)}$. Notice that this can only be done if the atomic samples are polarized oppositely along
the $x$-direction, so that the commutator $[\hat{J}_z^{(1)}+\hat{J}_z^{(2)},\hat{J}_y^{(1)}+\hat{J}_y^{(2)}]=0$.
Therefore, the first  light beam was used for creation of EPR-type entanglement, and another one for its
verification  through Eq. (\ref{sep_crit_Duan}).

Our aim here is to apply the QND atom-light interface to study entanglement generation with less restrictive conditions, i.e., we assume that:  (i) individual
magnetic field addressing of each atomic ensemble is not allowed and, (ii) the number of atomic ensembles can be made arbitrary.
Such experimental setups  that can be build, for instance, using optical microtraps \cite{birkl2007ApPhB..86..377L,PhysRevLett.89.220402} which allow for isotropic confinement of $10^{4}$ cold atoms, creating in this way mesoscopic atomic ensembles \footnote{In experiments with ultracold atoms one can reduce the number of atoms by $10^{4}$ to have the same opacity of the medium.}. In these setups, the preparation of each sample in a different initial magnetic state or the addressing of a sample with individual magnetic fields is out of reach. Despite these limitations, an array of microtraps offers considerable advantages, ranging from its experimental feasibility to possibility to  generate chains and arrays of atomic samples.
\begin{figure}
 a)\includegraphics[width=0.35\textwidth]{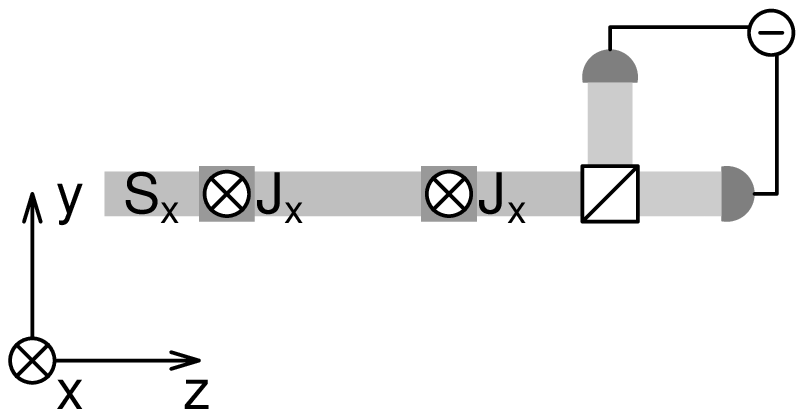}\\
 b)\includegraphics[width=0.35\textwidth]{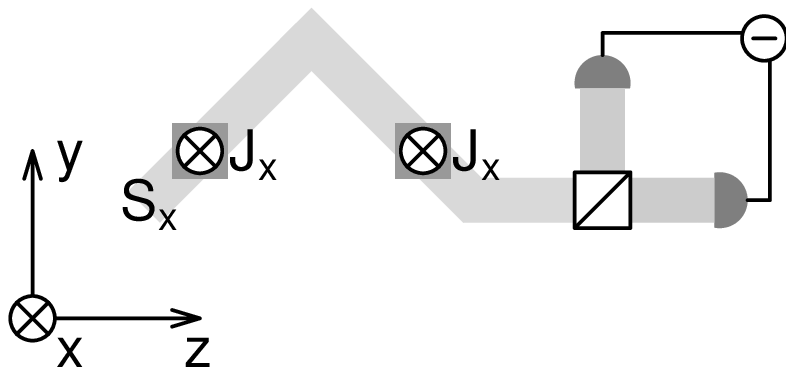}\\
 \caption{The simplest setup for generation and verification of bipartite entanglement between mesoscopic atomic ensembles.
a) First light pulse passing through the samples along direction $z$ entangles the samples. b) Second light pulse passing
through the samples at angles $\pi/4$ and $-\pi/4$, respectively, allows for verification of entanglement through a
variance inequality (see Eq. (\ref{sep_crit_Duan})).}\label{two_samples_z-sq}
\end{figure}

To better understand the dynamics of the interaction, we analyze in some detail the setup depicted in Fig. \ref{two_samples_z-sq}a. As indicated in Eq. (\ref{out_bipartite}), the light carries information about $\hat{J}_z^{(1)}+\hat{J}_z^{(2)}$ and the measurement of $\hat{S}_y^{\mathrm{out}}$ generates entanglement between the
atomic samples. Starting from the evolution equations and taking into account the light measurements, one can explicitly derive the variances of the atomic spin samples and interpret them in terms of squeezing.
In this view, the bipartite state of the ensembles is characterized by the following variances:
 \begin{eqnarray}\label{variances}
 \varsb{(\hat J_y^{(1)}+\hat J_y^{(2)})} &=&(1+2 \kappa^2) \hbar J_x, \label{Jyp}\\
 \varsb{(\hat J_y^{(1)}-\hat J_y^{(2)})} &=&\hbar J_x \label{Jym},\\
 \varsb{(\hat J_z^{(1)}+\hat J_z^{(2)})} &=&\frac{1}{1+2 \kappa^2} \hbar J_x \label{Jzp},\\
 \varsb{(\hat J_z^{(1)}-\hat J_z^{(2)})} &=&\hbar J_x \label{Jzm},
 \end{eqnarray}
where $\kappa=a \sqrt{S_x J_x}$.
The observables for which the separability criterion [Eq. (\ref{sep_crit_Duan})] is violated
correspond to $J_z^{(1)}+J_z^{(2)}$ and $J_y^{(1)}-J_y^{(2)}$ . Such a measurement induces
squeezing on the variances along the $z$-direction below the vacuum limit, as clearly indicated by Eq. (\ref{Jzp}).

The verification of entanglement involves measurement of the sum of the variances corresponding to Eqs. (\ref{Jym}) and (\ref{Jzp}).
In order to do this with a single beam we use light propagating at different angles, as schematically depicted in Fig. \ref{two_samples_z-sq}b.
In this case, according to Eq. ($\ref{Sy}$) we obtain
\begin{equation}\label{out_measur_bipart}
\hat{S}_y^{\mathrm{out}}=\hat{S}_y^{\mathrm{in}}-\frac{a}{\sqrt{2}}S_x\left[ \left(\hat{J}_z^{(1)}+\hat{J}_z^{(2)})+(\hat{J}_y^{(1)}-\hat{J}_y^{(2)}\right)\right].
\end{equation}
Since within this scheme
$\meanv{\hat{J}_y^{(i)}\hat{J}_z^{(j)}}=\meanv{\hat{J}_y^{(i)}}\meanv{\hat{J}_z^{(j)}}$, the variance of the output can be written as
\begin{eqnarray}\label{out_measur_bipart_var}
& &\var{\hat{S}_y^{\mathrm{out}}}=\var{\hat{S}_y^{\mathrm{in}}}\nonumber\\
&+&\frac{a^2}{2}S_x^2\left\{
\varsb{(\hat{J}_z^{(1)}+\hat{J}_z^{(2)})}+\varsb{(\hat{J}_y^{(1)}-\hat{J}_y^{(2)})}\right\}.\nonumber\\
\end{eqnarray}
For details concerning the experimental measurement of such variances   the reader is referred to \cite{julsgaardPHD2003,shersonARX}.
This shows that entanglement between two identically polarized atomic ensembles can be generated and verified using only two beams and no additional magnetic fields, if the second field impinges on the two samples at certain angles.

\begin{figure}
 a)\includegraphics[width=0.35\textwidth]{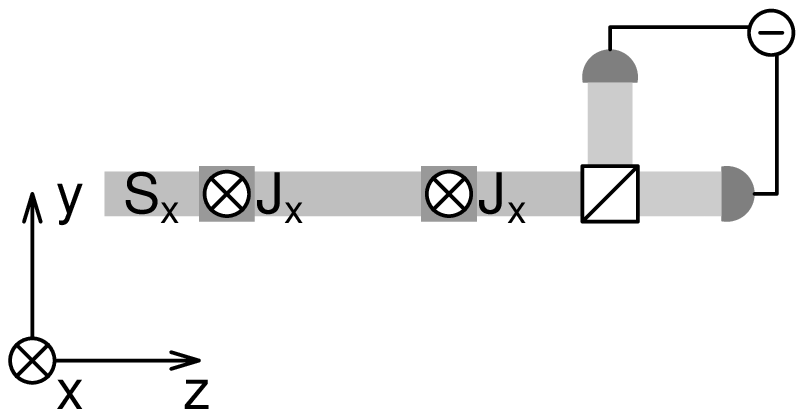}\\
 b)\includegraphics[width=0.31\textwidth]{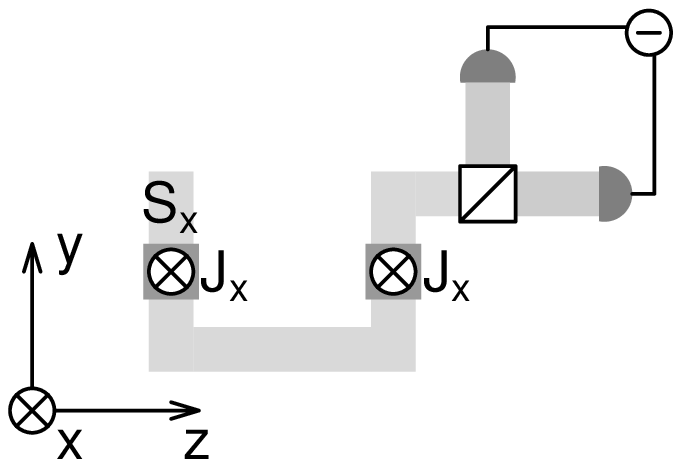}\\
 c)\includegraphics[width=0.35\textwidth]{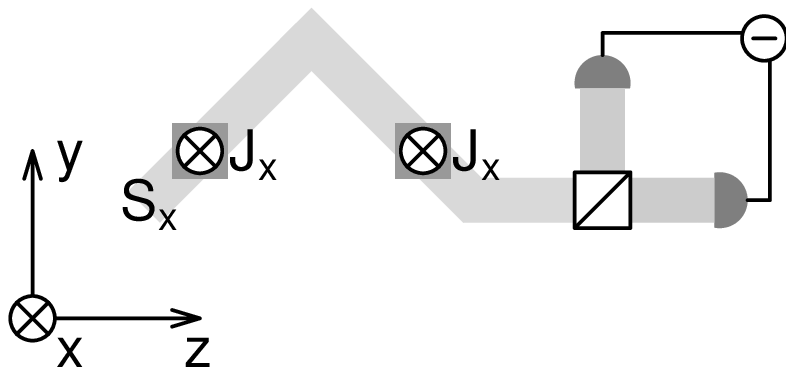}\\
 \caption{The setup for generation and verification of bipartite entanglement between atomic ensembles in which squeezing
is introduced in two variables a) $J_z^{(1)}+J_z^{(2)}$ and b) $J_y^{(1)}-J_y^{(2)}$. The third pulse depicted in figure
c) allows for verification of entanglement through variance inequality. It should be emphasized that the first and last step are exactly the same as in Fig. \ref{two_samples_z-sq}.}\label{two_samples_z-sq_and_y-sq}
\end{figure}

To increase entanglement between the two samples one should introduce global squeezing in two independent variables.
This is schematically depicted in Fig. \ref{two_samples_z-sq_and_y-sq}a and \ref{two_samples_z-sq_and_y-sq}b. The first beam introduces squeezing in $\hat{J}_z^{(1)}+\hat{J}_z^{(2)}$ variable. Then, a second beam propagating through the first sample at an angle $\alpha=\pi/2$ and through the second one at an angle $\alpha=-\pi/2$ generates squeezing in $\hat{J}_y^{(1)}-\hat{J}_y^{(2)}$.
Note that these are commuting operators, so the second beam would not
change the effect of the first one (squeezing of $J_z^{(1)}+J_z^{(2)}$). Within this scheme
one reproduces the results of Julsgaard {\it et al} \cite{2001Natur.413..400J} without individual addressing. The verification of entanglement (see Fig. \ref{two_samples_z-sq_and_y-sq}c) can be done as  previously described.

Interesting enough, this geometrical approach also opens the possibility of deleting all the entanglement created by the first light beam, if intensities are appropriately adjusted. The entanglement
procedure is intrinsically irreversible because of the projective measurement, so coming deterministically back to the initial state is not a trivial task.
In \cite{filip,filipPRA2003}, a quantum erasing  scheme
in  Continuous Variables systems was proposed. The measurement of the meter coordinate entangled with
the quantum system leads to a backaction on it.
The authors shown that it is possible to erase the action of the measurement and restore the the original state of the system.
Here we are interested in deleting the measurement induced entanglement between two atomic samples,
exploiting the squeezing and antisqueezing effects produced by the laser beams.
\begin{figure}
 a)\includegraphics[width=0.35\textwidth]{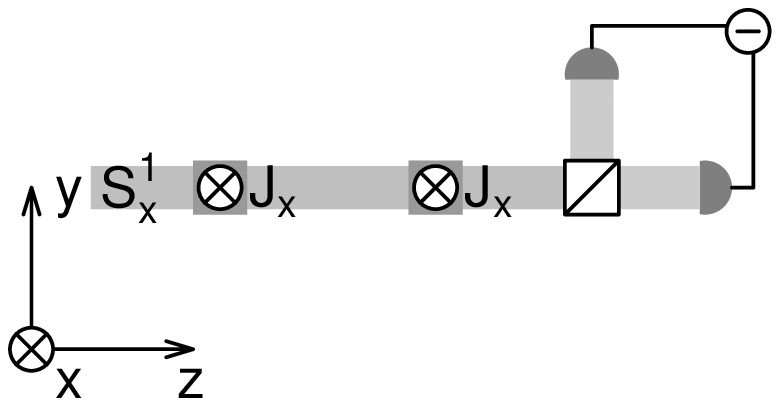}\\
 b)\includegraphics[width=0.33\textwidth]{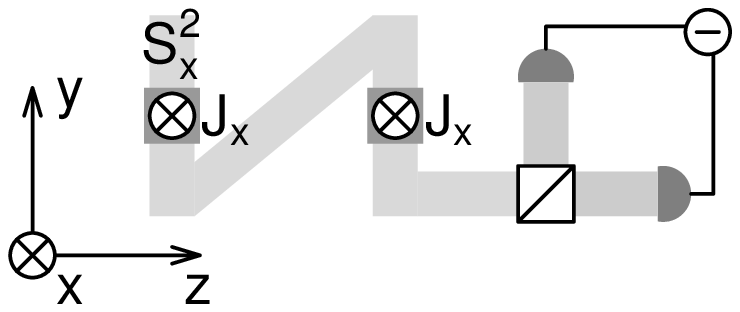}\\
 \caption{Entanglement eraser scheme realized by two pulses of different intensity, $\kappa_1^2 \propto N_{\mathrm{ph}}^{(1)}$ and $\kappa_2^2 \propto N_{\mathrm{ph}}^{(2)}$. See the text for details.}\label{two_samples_eraser}
\end{figure}
Let us assume that the first entangling beam, characterized by a coupling constant $\kappa_1^2 \propto N_{\mathrm{ph}}^{(1)}$, propagates along the $z$-direction, exactly as it was described before (see Fig. \ref{two_samples_eraser}a). The interaction, followed by the measurement of light, creates
squeezing in the observable $\hat{J}_z^{(1)}+\hat{J}_z^{(2)}$ accompanied by anti-squeezing in the conjugate variable $\hat{J}_y^{(1)}+\hat{J}_y^{(2)}$ [Eqs. (\ref{Jyp}) and (\ref{Jzp})].  Assume a second beam characterized by a coupling constant $\kappa_2^2 \propto N_{\mathrm{ph}}^{(2)}$ propagates through the samples in an orthogonal direction with respect to the first beam as shown in Fig. \ref{two_samples_eraser}b. This corresponds to setting $\alpha=\pi/2$ in the Hamiltonian of Eq. (\ref{ham_alpha}). In this setup the measurement of the variable $\hat{S}_y^{\mathrm{out}}$ introduces squeezing in the conjugate variable $\hat{J}_y^{(1)}+\hat{J}_y^{(2)}$.

The bipartite state created by propagation and measurement of the  first and second beam is characterized by the variances

\begin{eqnarray}\label{vars_sec_beam}
&&\varsb{(J_y^{(1)}+J_y^{(2)})}=\frac{2 \kappa _1^2+1}{\left(4 \kappa _1^2+2\right) \kappa _2^2+1}\hbar J_x\\
&&\varsb{(J_y^{(1)}-J_y^{(2)})}=\hbar J_x\\
&&\varsb{(J_z^{(1)}+J_z^{(2)})}=\left(2 \kappa _2^2+\frac{1}{2 \kappa _1^2+1}\right)\hbar J_x\\
&&\varsb{(J_z^{(1)}-J_z^{(2)})}=\hbar J_x.
\end{eqnarray}
A close look at these equations shows that the second beam can lower or even completely destroy entanglement between the samples.
This happens when
\begin{equation}
\kappa_2^2=\frac{\kappa _1^2}{2\kappa_1^2+1}.
\end{equation}
In such case the atomic ensembles are left in a vacuum (uncorrelated) state, however, displaced. Hence, the overall effect of these two beams is simply a displacement of the initial vacuum state. The value of the displacement depends on the coupling constant $\kappa_1 $ and outputs obtained in the measurement of the light
polarization component, $S_y$, of both beams. Therefore,  it will vary run to run.

\begin{figure}
\includegraphics[width=0.47\textwidth]{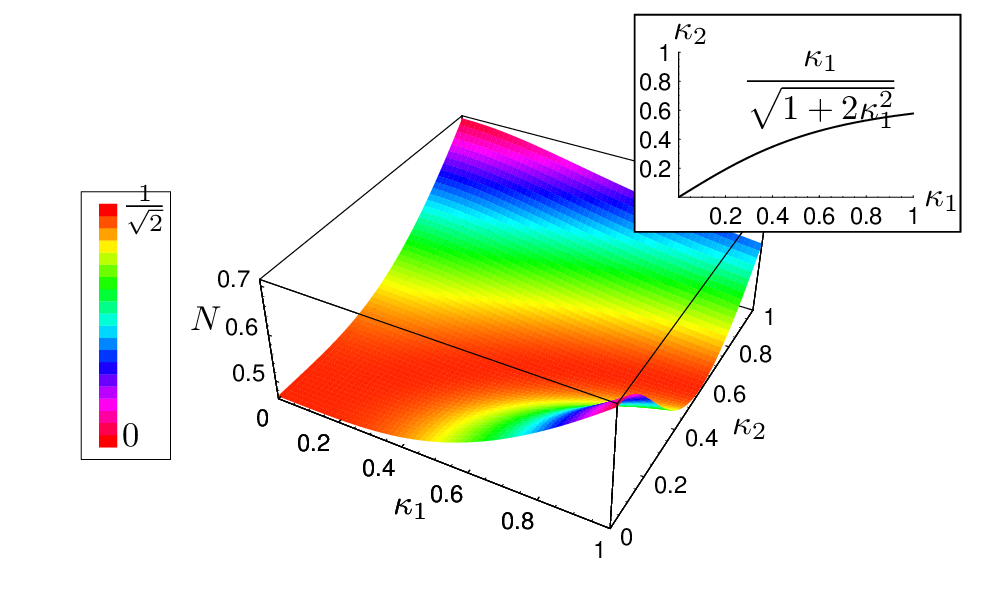}\\
 \caption{(Color online).The negativity of a bipartite state of atomic ensembles after passage and measurement of two beams
of coupling parameters $\kappa_1$ and $\kappa_2$ (see Fig. \ref{two_samples_eraser}). For specific values of $\kappa_1$ and $\kappa_2$ the negativity approaches zero  (see the inset).}\label{negativity2}
\end{figure}

Using negativity as an entanglement measure \cite{vidalPRA2002}, computed by the symplectic eigenvalues
of the partial time reversal of  covariance matrix,
one finds that indeed entanglement diminishes continuously or even disappears depending on the value of $\kappa_2$, as shown in Fig. \ref{negativity2}.

%
Notice that for every fixed value of $\kappa_1$ there always exists a value of $\kappa_2$ for which negativity becomes zero and the state becomes separable even though it was entangled after interaction and measurement of the first beam.


\section{Multipartite entanglement}\label{multi}\label{IV}

In what follows we generalize our study to the multipartite scenario and we present different strategies to achieve multipartite entanglement without individual addressing. The strategies will not depend on the total number of samples but only if this number is odd or even.  For the verification part,
we shall adopt the criteria for multipartite entanglement, expressed via inequalities for variances of quadratures, derived by
van Loock and Furusawa \cite{vanloock-PhysRevA.67.052315}. We rewrite the inequalities for angular momentum variables as follows.
If an $N_s$-mode state $\varrho$ is separable, then the sum of variances of the following operators:
\begin{eqnarray}
\hat{u}&=&h_1 \hat{J}_y^{(1)}+\ldots +h_{N_s} \hat{J}_y^{(N_s)}\nonumber\\
\hat{v}&=&g_1 \hat{J}_z^{(1)}+\ldots +g_{N_s} \hat{J}_z^{(N_s)}
\end{eqnarray}
is bounded from above by a
function of the coefficients $h_1,\ldots,h_{N_s},g_1,\ldots,g_{N_s}$ and $J_x$.
Mathematically the inequality is expressed as
\begin{equation}\label{sep_crit_vLF}
\var{\hat{u}}+\var{\hat{v}}\geq f(h_1,\ldots,h_{N_s},g_1,\ldots,g_{N_s})\hbar J_x
\end{equation}
with
\begin{eqnarray}
&&\hspace{-12pt}f(h_1,\ldots,h_{N_s},g_1,\ldots,g_{N_s})\\
&&\hspace{-12pt}=\big|h_m g_m+\sum_{r\in I} h_{r}g_{r}\big|+\big|h_n g_n+\sum_{s\in I^{'}} h_{s}g_{s}\big|.\nonumber
\end{eqnarray}
In the above formula two modes, $m$ and $n$, are distinguished and the remaining modes are grouped in two disjoint sets $I$ and $I^{'}$. The criterion (\ref{sep_crit_vLF}) holds for all bipartite splittings of a state defined by the sets of indices $\{m\}\cup I$ and $\{n\}\cup I^{'}$, and for every choice of parameters $h_1,\ldots,h_{N_s},g_1,\ldots,g_{N_s}$.
For example, in case of three samples we have
$f(h_1,h_2,h_3,g_1,g_2,g_3)=(|h_n g_n|+|h_k g_k+h_m g_m|)$,
where $(n,m,k)$ is some permutation of the sequence $(1,2,3)$, and the coefficients $h_1,h_2,h_{3},g_1,g_2,g_{3}$ are arbitrary real numbers.

\subsection{GHZ-like states}
Genuine multipartite entanglement between any number of equally polarized atomic modes can
be obtained with a single beam propagating through all of them followed by projective measurement of the light.
After the measurement, the $N_s$-mode variable $\hat{J}_z^{(1)}+\ldots+\hat{J}_z^{(N_s)}$ is squeezed.
This is a trivial extension of the bipartite scheme schematically shown in Fig. \ref{two_samples_z-sq}a.

The phenomenon of destruction of entanglement by squeezing of the conjugate variable, which was discussed in the previous section for two modes, can be also found in the multimode setup. The entanglement prepared with the light beam characterized by the coupling constant $\kappa_1$ can be erased by the second orthogonal beam with appropriately adjusted intensity.
The relation between the coupling constants for which entanglement is removed from the system is
\begin{equation}
\kappa_2^2=\frac{\kappa_1^2}{1+N_s \kappa_1^2}.
\end{equation}
One can see that with increasing number of samples the value of $\kappa_2$ required to delete entanglement decreases.

To generate a maximally entangled GHZ state with $N_s$-parities, simultaneous squeezing in more independent variables is needed. By independent here we mean commuting linear combinations of atomic spin operators.
The most straightforward way to do it is to generate squeezing in the variable $\hat{J}_z^{(1)}+\ldots+\hat{J}_z^{(N_s)}$ and in the pairwise differences of angular momenta: $\hat{J}_y^{(i)}-\hat{J}_y^{(j)}$ ($1 \leq i,j\leq N_s$, $i\neq j$) (see \cite{PhysRevLett.84.3482,Braunstein:513}). An entangled state with such properties can be realized by generalization of the bipartite scheme summarized in Figs. \ref{two_samples_z-sq_and_y-sq}a and \ref{two_samples_z-sq_and_y-sq}b. Notice, however, that the last step should be repeated for all combinations of $i>j$. The final variances characterizing the state would be
%
\begin{eqnarray}
\varsb{(\hat{J}_z^{(1)}+\ldots+\hat{J}_z^{(N_s)})}=\frac{N_s}{2+2 N_s \kappa^2} \hbar J_x\\
\varsb{(\hat{J}_y^{(i)}-\hat{J}_y^{(j)})}=\frac{1}{1+N_s \kappa^2} \hbar J_x \quad (i\neq j).
\end{eqnarray}
Thus the samples are in a genuine $N_s$-mode GHZ state. Within this scheme the number of measurements, one has to perform in order to create entanglement, grows quadratically with the number of samples. Also verification implies checking all the inequalities of the type
\begin{eqnarray}\label{pairwise}
&& \varsb{(\hat{J}_y^{(i)}-\hat{J}_y^{(j)})}+\varsb{(\hat{J}_z^{(1)}+\ldots+\hat{J}_z^{(N_s)})} \\
&&\geq 2 \hbar J_x   \; (i>j).\nonumber
\end{eqnarray}

While the above procedure works for an arbitrary number of samples, to optimize it we consider separately even and odd $N_s$.

For even number of ensembles $N_s=2M$ the optimal approach generalizes the one proposed for two samples and summarized in Figs. \ref{two_samples_z-sq_and_y-sq}a and \ref{two_samples_z-sq_and_y-sq}b.
In first step we generate squeezing in $\hat{J}_z^{(1)}+\ldots+\hat{J}_z^{(2M)}$. As the second step we squeeze the observable $\hat{J}_y^{(1)}-\hat{J}_y^{(2)}+\ldots+(-1)^{2M-1} \hat{J}_y^{(2M)}$ with the second beam passing through the $i$th sample at an angle $(-1)^{i-1}\pi/2$.
The final state is pure and genuine multipartite entangled. The entanglement can be detected using the criterion (\ref{sep_crit_vLF}) with the two squeezed observables discussed in this paragraph. The measurement of light propagating through the $i$th sample at an angle $(-1)^{i-1}\pi/4$ gives at the level of variances
\begin{eqnarray}
&&\var{\hat{S}_y^{\mathrm{out}}}=\var{\hat{S}_y^{\mathrm{in}}}\nonumber\\
&&+\frac{a^2}{2}S_x^2\varsb{(\hat{J}_y^{(1)}-\hat{J}_y^{(2)}+\ldots+(-1)^{2M-1} \hat{J}_z^{(2M)})}\nonumber\\
&&+\frac{a^2}{2}S_x^2\varsb{(\hat{J}_z^{(1)}+\ldots+\hat{J}_z^{(2M)})}.
\end{eqnarray}
Therefore, again a single beam can be used for verification of entanglement. The same criterion and the above measurement scheme can be applied not only to detect the entanglement in the above setup but also in those proposed before, i.e., (i) the state with squeezing only in $J_z^{(1)}+\ldots+J_z^{(2M)}$ (after interaction and measurement of only the first beam), and (ii) the state with squeezing in $J_z^{(1)}+\ldots+J_z^{(2M)}$ and all combinations $J_y^{(k)}-J_y^{(l)}$ $(k\neq l)$. The reduction in the number of measurements is significant. Moreover, a recently proposed multi-pass technique \cite{namiki2009} could lead to simplification of geometry.

Optimization of the scheme for odd number of atomic ensembles within this geometric approach is to our knowledge not possible. Even though it is possible to find independent variables involving all the samples, it is not clear what  geometry should be applied in order to measure these operators.

A different way to deal with multimode entanglement of odd number of samples is to generalize directly the bipartite scheme of Julsgaard {\it et al.}, i.e., polarize the samples in such a way that the collective polarization $\sum_i J_x^{(i)}$  is zero. Moreover, each sample should experience a different local magnetic field. In such system it is possible to generate squeezing in appropriately redefined (due to Larmor precession) operators $\sum_i \hat{J}_y^{(i)}$ and $\sum_i \hat{J}_z^{(i)}$, using a single light beam. This is possible due to the choice of the initial polarization of the samples making the redefined operators to commute. Analogously to the bipartite case an entanglement test that can be applied involves measurement of variances of the sums of angular momentum components and reads
\begin{equation}
\var{\sum_{i}\hat{J}_y^{(i)}}+\var{\sum_{i}\hat{J}_z^{(i)}}\geq N_s \hbar J_x.
\end{equation}
\subsection{Cluster-like states}
The analyzed setup allows for generation of Continuous Variables cluster-like states \cite{vanloockPRA2007}
We associate the modes of  the N-mode system with the vertices of a graph $G$. The edges between the
vertices define the notion of nearest neighborhood. By $N_a$ we denote the set of nearest neighbors  of vertex $a$.
A cluster is a connected graph.
For angular momentum variables,  cluster states are defined only asymptotically as those with infinite squeezing in the variables
\begin{equation}\label{eqn:clustervariables}
\hat{J}^{(a)}_z-\sum_{b\in N_a } \hat{J}^{(b)}_y
\end{equation}
for all $a \in G$.
Cluster-like states are defined when  the squeezing is finite.

Given a set of atomic ensembles, it is possible to  create a chosen cluster-like state by squeezing the required combinations of variables (\ref{eqn:clustervariables}). Since they commute, it is possible to squeeze them sequentially.
Hereafter, we will illustrate the procedure by a simple example. The method is general and can be applied to create any clusterlike state.
\begin{figure}
a)\includegraphics[width=0.15\textwidth]{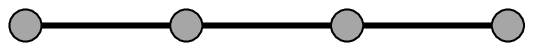}\\
b)\includegraphics[width=0.35\textwidth]{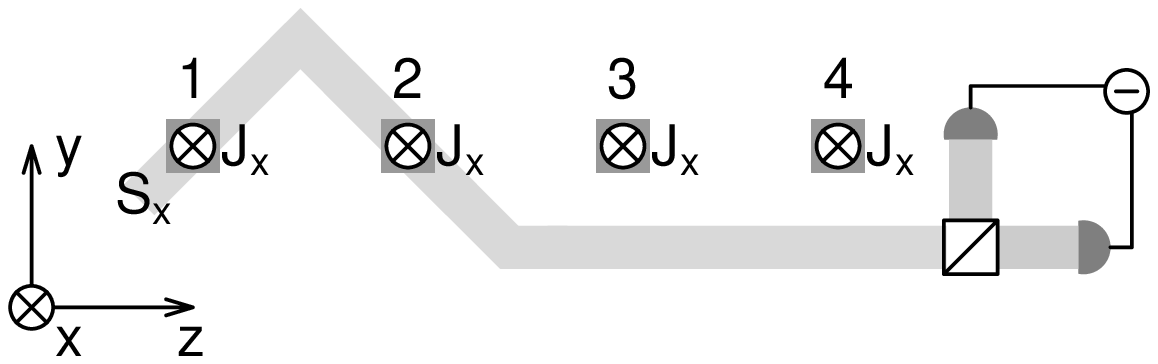}\\
c)\includegraphics[width=0.35\textwidth]{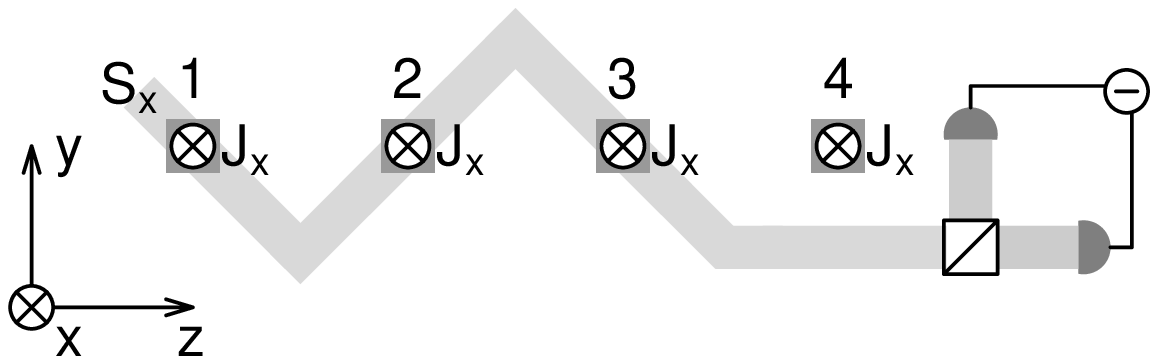}\\
d)\includegraphics[width=0.35\textwidth]{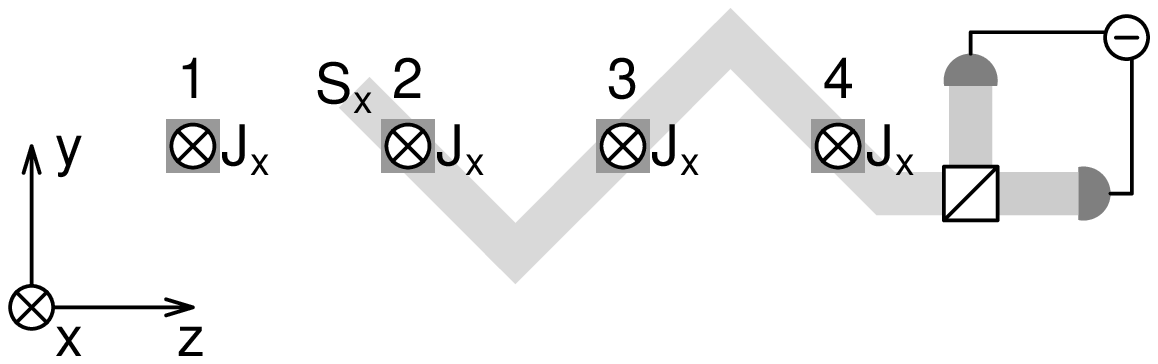}\\
e)\includegraphics[width=0.35\textwidth]{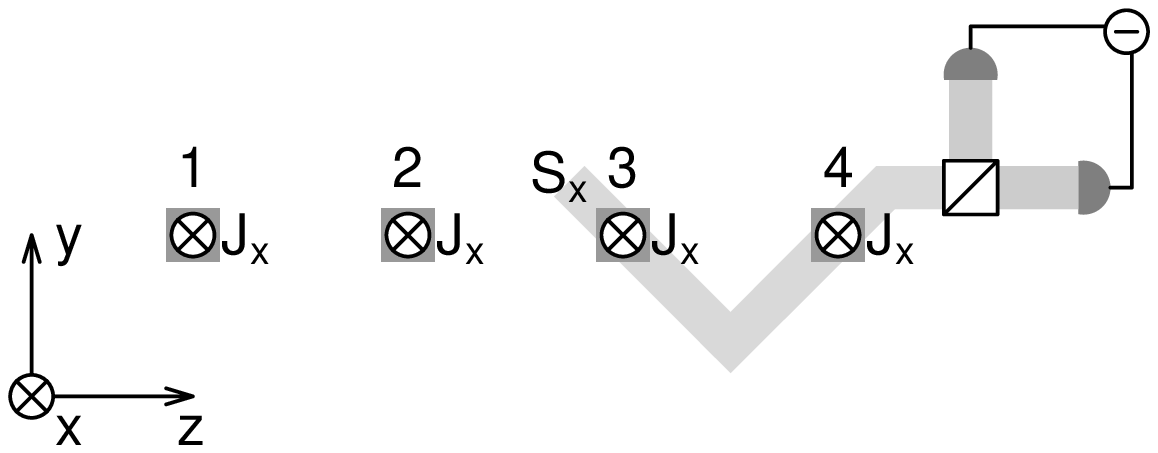}
 \caption{Generation of the cluster state schematically depicted in a).The sequence of beams squeeze the following
variables: b) $\hat{J}^{(1)}_z{}'-\hat{J}^{(2)}_y{}'$,  c) $\hat{J}^{(2)}_z{}'-\hat{J}^{(1)}_y{}'-\hat{J}^{(3)}_y{}'$,
d) $\hat{J}^{(3)}_z{}'-\hat{J}^{(2)}_y{}'-\hat{J}^{(4)}_y{}'$, e) $\hat{J}^{(4)}_z{}'-\hat{J}^{(3)}_y{}'$
}\label{fig:cluster}
\end{figure}
In Fig. \ref{fig:cluster} we show how to create the simplest four-site (linear) cluster state.
Let us introduce the new variables for each sample
\begin{eqnarray}
\hat{J}^{(i)}_y{}'&=&\frac{1}{\sqrt{2}}\left( \hat{J}^{(i)}_y-\hat{J}^{(i)}_z\right) \nonumber\\
\hat{J}^{(i)}_z{}'&=&\frac{1}{\sqrt{2}}\left( \hat{J}^{(i)}_y+\hat{J}^{(i)}_z\right).
\end{eqnarray}
The squeezing in the combinations of the new variables is produced by passing light as depicted in Fig.\ref{fig:cluster}b-\ref{fig:cluster}e. For example the squeezing in $\hat{J}^{(1)}_z{}'-\hat{J}^{(2)}_y{}'$
is generated when  light  passes  only through samples $1$ and $2$ at angles $\pm \pi/4$ respectively (see Fig. \ref{fig:cluster}b).
All the other required combinations are squeezed in a similar way.

In order to verify that the state is entangled it is enough to check the set of variance
inequalities given in \cite{yukawaPRA2008}.  This can be done, for example, by repetition of each step as
first proposed in \cite{2001Natur.413..400J,julsgaardPHD2003}.


\section{Summary}\label{V}

Summarizing, we have studied multipartite mesoscopic entanglement using a quantum atom-light interface in various physical setups, in particular those in which the ensembles cannot be addressed individually. Exploiting a geometric approach in which light beams propagate through the atomic samples at different angles makes it possible to establish and verify EPR-bipartite entanglement and GHZ-multipartite entanglement with a minimal number of light passages and measurements, so that the quantum nondemolition character of the interface is preserved.
We have also shown how  to  generate cluster-like states  by a similar technique.

Furthermore, we have shown that  the multipartite entanglement created by the quantum interface of a single light beam can be appropriately tailored and even completely erased by the action of a second pulse with different intensity. This control widens the possibilities offered by measurement induced entanglement to perform quantum information tasks.


\section*{Acknowledgments}

The authors acknowledge support from MEC (Spain) Grant No. FIS2008-01236, Generalitat de Catalunya Grant No. 2005SGR-00343 and EU IP SCALA. J.S. acknowledges the financial support from the ``Universitat Aut\`{o}noma de Barcelona''. Useful comments from Maciej Lewenstein and Remigiusz Augusiak are also acknowledged.

\bibliographystyle{apsrev}
\bibliography{/media/disk/universita/bibliografia/bibl-cont_var,/media/disk/universita/bibliografia/bibl-iontraps,Simon}

\end{document}